\documentclass{aa}  
\usepackage{graphics}

\begin{document}

\def\gtapp
{\mathrel{\hbox{\raise0.3ex\hbox{$>$}\kern-0.8em\lower0.8ex\hbox{$\sim$}}}}
\def\ltapp
{\mathrel{\hbox{\raise0.3ex\hbox{$<$}\kern-0.75em\lower0.8ex\hbox{$\sim$}}}}

\title{Mid-Infrared observations of NGC\,1068 with the Infrared Space   
Observatory\thanks{Based on observations with the ISO satellite, an
ESA project with instruments funded by ESA Member States (especially
the PI countries: France, Germany, the Netherlands and the United
Kingdom) and with the participation of ISAS and NASA.}}
  
\author{E.~Le~Floc'h \inst{1,2}  
\and I.F.~Mirabel \inst{1,3}
\and O.~Laurent \inst{1,4}  
\and V.~Charmandaris \inst{5}  
\and P.~Gallais \inst{1}  
\and M.~Sauvage \inst{1}  
\and L.~Vigroux \inst{1}  
\and C.~Cesarsky \inst{6}  
}
  
\institute{CEA/DSM/DAPNIA Service d'Astrophysique, F-91191 Gif-sur-Yvette,  
France  
\and  
European Southern Observatory,   
Alonso de Cordova 3107, Casilla 19001, Santiago, Chile  
\and  
Instituto de Astronom\'\i a y F\'\i sica del Espacio, cc 67, suc 28.   
1428 Buenos Aires, Argentina
\and   
Max Planck Institut f\"ur extraterrestrische Physik, Postfach 1312, 85741  
Garching, Germany  
\and  
Cornell University, Astronomy Department, Ithaca, NY 14853, USA  
\and   
European Southern Observatory, Karl-Schwarzschild-Str, D-85748   
Garching bei M\"unchen, Germany} 
  
\titlerunning{Mid-infrared observations of NGC\,1068 with ISOCAM}
\authorrunning{Le Floc'h et al.}
  
\offprints{E. Le Floc'h (elefloch@cea.fr)}  
  
\date{Received 25 October 2000 / Accepted 22 December 2000}  
  
\abstract{    
We report on Mid-Infrared (MIR) observations of the Seyfert\,2 galaxy
NGC\,1068, obtained with ISOCAM in low-resolution spectro-imaging
mode. The spatial resolution ($\sim$\,5\,{\arcsec}) allows us to
disentangle the circumnuclear starburst regions from the emission of
the active galactic nucleus (AGN). The global spatial
distribution of the Unidentified Infrared Bands (UIBs) is similar to
the cold dust component, traced by the 450\,$\mu$m emission and the
gaseous component obtained from the {$^{12}$CO(1-0)}
map. However, a shift between the maximum of the UIB and
450\,$\mu$m emission is clearly seen in our maps. The UIBs in the MIR
(5-16\,$\mu$m) originate almost exclusively from the starburst regions
in the galactic disk with an emission peaking at the extremity of the
stellar/gaseous bar at a distance of 1\,kpc from the AGN.  The
spectrum of the nucleus is characterized {\em over the whole
5-16\,$\mu$m range} by a strong continuum which can be fitted with a
power law of index $\alpha=-1.7$. Moreover, the high [NeIII]/[NeII]
ratio ($\gtapp 2.5$) in the nuclear region argues for a hard radiation
field from the AGN.  Observations indicate that the AGN in NGC\,1068
contributes less than $\sim$5\% to the total integrated UIB emission
even though its hot dust continuum contributes as much as 75\% to the
total MIR flux. On the contrary, the nuclear contribution to
the cold dust emission decreases considerably at submillimeter
wavelengths and does not represent more than 25\% of the total
integrated emission at 450\,$\mu$m.
}  

\maketitle
  
\section{Introduction}  
  
Due to its proximity (D=14.4\,Mpc, 1\,{\arcsec} on the sky corresponds to a
physical separation of only 72\,pc, \cite{Tully}), NGC\,1068 has
become one of the best known active galaxies. It harbors the closest
Seyfert\,2 nucleus, it has been extensively observed from UV to radio
wavelengths, and therefore it is regarded as a favorite target for
high resolution observations. As a result, NGC\,1068 is now the
prototypical galaxy where one can test both models of emission due the
presence of a central black hole (especially in the MIR; see
\cite{Pier92}, \cite{Efstathiou}, \cite{Granato}), as well as models 
describing the kinematics of circumnuclear starbursts associated with
a bar, which are often encountered around active nuclei
(e.g. \cite{Eva} and references therein).

The presence of a massive black hole in the central region of this
galaxy is supported by the observation of a radio jet which originates
from an unresolved region of 60\,mas (\cite{Muxlow}). Moreover, X-ray
observations of the central region suggest large obscuration
(e.g. \cite{Iwasawa}, \cite{Matt}) presumably due to a thick 
torus invoked in the unified scheme (Antonucci 1993). Evidence for
such a dusty torus and obscuring material around the nucleus have been
recently reported in the Near-Infrared (NIR) using adaptive optics
(e.g. \cite{Rouan}) and in the radio using VLBA observations
(\cite{Gallimore} 1997).  The precise localization of the central
black hole is now accurately defined, in particular by the spatial
coincidence of 1) the peak of the 12.4\,$\mu$m intensity
(\cite{Braatz}), 2) the OH and H$_2$O maser emission (\cite{Gallimore}
1996), 3) the center of the UV-optical polarization map
(\cite{Capetti}), and 4) the apex of the conical shape of the narrow
line region (NLR) observed by HST (\cite{Macchetto}).

Besides the contribution of the Seyfert nucleus, an important fraction
of the bolometric luminosity of NGC\,1068
($L_{bol}=3\times10^{11}L_{\sun}$, \cite{Telesco84} 1984) is also attributed
to a circumnuclear star-forming ring of 3\,kpc in diameter
(\cite{Telesco} 1988).  Using the CO rotational line transitions as tracers
of the cold molecular gas, the ring can be resolved into two distinct
spiral arms (\cite{Helfer}) along which a string of several massive
and clumpy HII regions is observed (\cite{BHSa},
\cite{BHSb}).  These spiral arms originate from the ends of a gaseous
and stellar bar observed at NIR wavelengths (\cite{Scoville},
\cite{Thronson}).  The same overall structure can also be seen in the
cold/dense molecular gas via the HCN(1-0) emission (\cite{Tacconi}) as
well as in more recent high resolution CO imaging of the galaxy
(\cite{Eva}). A study of the kinematics of the gas suggests
that the molecular gas motions are explained in terms of Lindblad
resonances associated with a barred potential, in combination with a
warp of the gaseous disk in the ring (\cite{Eva}).  The bar as well as
other gaseous features associated with cold dust emission in the
spiral arms are also revealed in the Far-Infrared (FIR;
\cite{Papadopoulos99}).

MIR observations appear to be particularly enlightening in studying
NGC\,1068 since the UV photons from the starburst and AGN
activity are absorbed by dust and re-emitted in the MIR. 
Moreover, the lower extinction (A$_{15\,\mu
m}$$\sim$\,A$_{V}$/70, \cite{Mathis}) allows us to probe deeper into
the regions which are obscured by large amounts of dust.  After
describing the ISO observations and data reduction in
section~\ref{sec:red}, we present, in section~\ref{sec:analysis}, the
details on the MIR morphology of the circumnuclear environment as well
as the AGN/starburst spectral properties. We discuss the impact of
these new data in section~\ref{sec:discuss}.

\section{Observations and data reduction}  
\label{sec:red}  

\begin{figure}  
\resizebox{\hsize}{!}{\includegraphics{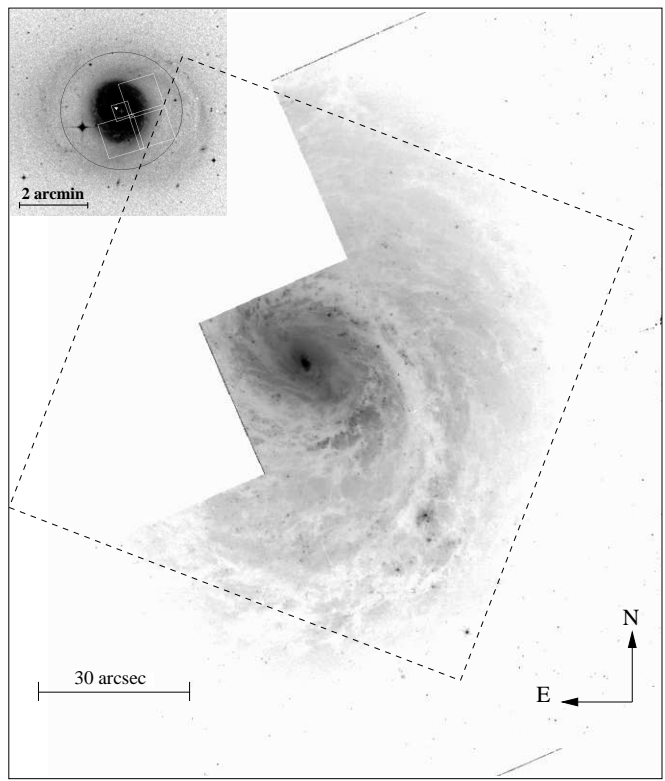}}  
  \caption{An optical V-band image of the central region of NGC\,1068 
   taken with HST.
   The field of view observed by ISOCAM is marked with the dashed square.    
   One should note that even though it is not obvious from the above low
   contrast picture, the disk of the galaxy extends to 7.1 by 6.0 arcmin.
   The inset at the upper right corner shows a Digitized Sky Survey image
   of the same band with the footprint of the HST/WFPC2 marked on it.}
  \label{fig:map_view}  
\end{figure}  
  
NGC\,1068 was observed with the ISOCAM camera (\cite{Cesarsky})
on-board the Infrared Space Observatory (\cite{Kessler}).  The data
were obtained in low-resolution spectro-imaging mode, using the
Continuously Variable Filter (CVF), resulting in a full coverage of
the 5.1--16.3\,$\mu$m wavelength range with a spectral resolution
between 30 and 40. The pixel size is 3\,{\arcsec} providing a total
field of view of 96\,{\arcsec}$\times$96\,{\arcsec} (see
Figure~\ref{fig:map_view}) with a full-width at half-maximum (FWHM) 
for the point spread function (PSF) between
4\,{\arcsec} and 6\,{\arcsec} from 5.2 to 16.3\,$\mu$m. The data
reduction and analysis were performed using the CAM Interactive
Analysis (CIA\footnote{CIA is a joint development by the ESA
astrophysics division and the ISOCAM consortium.})
following the standard techniques described in \cite{Starck} which
include the correction for the dark current, the cosmic-rays and the
memory effect of the detector. We also corrected the frames from the
jitter effects which are due to the combination of 1) the satellite
motion, 2) the continuous translation of the source on the detector as
a function of the observed wavelength, which is an intrinsic
feature of the ISOCAM CVF, and 3) the discontinuous shift of
the target when changing from one CVF sector to another. This
correction was performed by translating each resampled frame with
respect to a common astrometric reference.

\begin{figure*}[!t]
\vspace{-0.5cm}
\resizebox{\hsize}{!}{\includegraphics{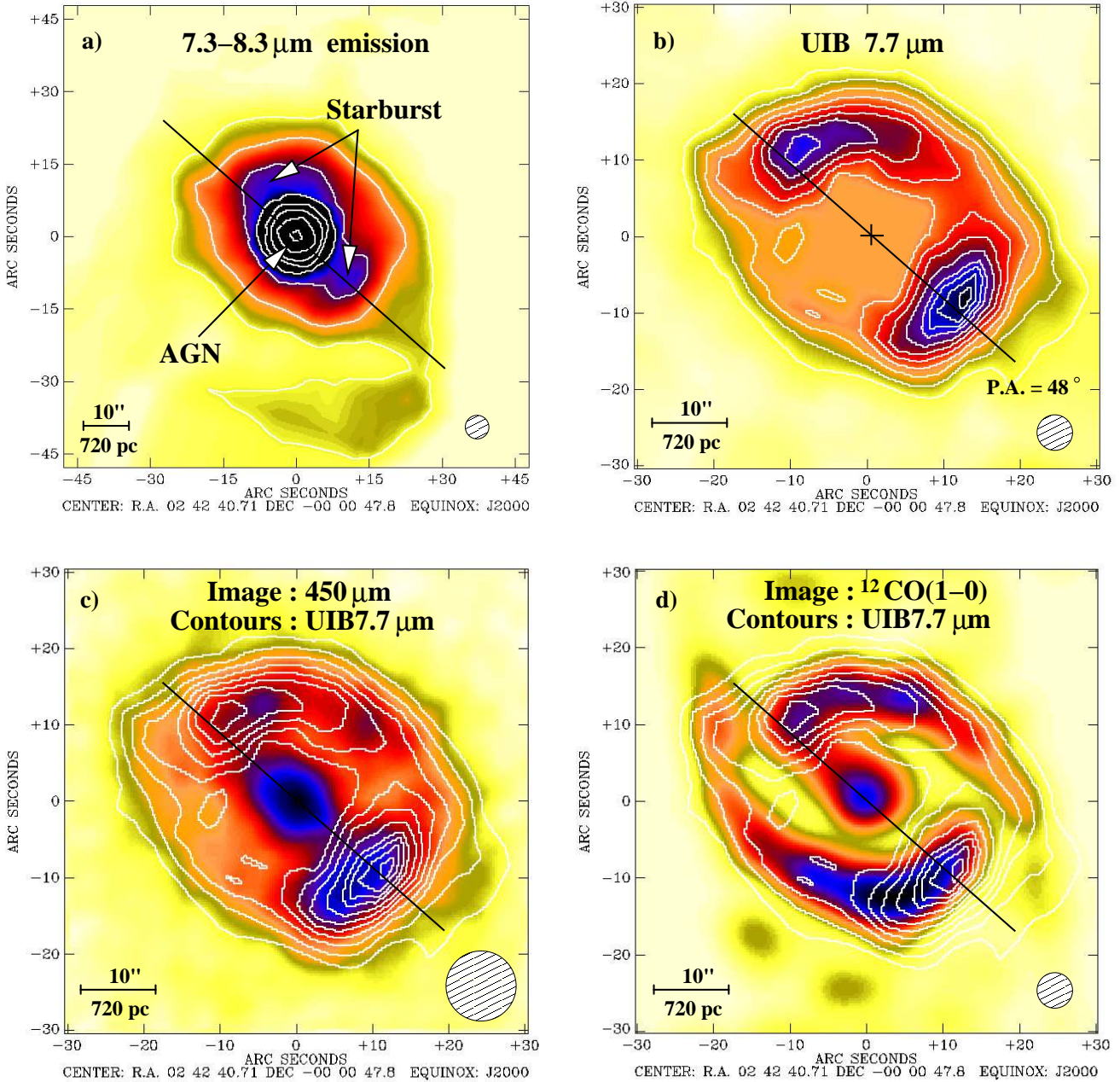}}  
\vspace{-0.9cm}
  \caption{a) A Mid-Infrared image of NGC\,1068 covering the
  wavelength range of 7.3 -- 8.3\,$\mu$m with exponential contours at
  the same wavelength. North is up and east is to the left. 
  The contour levels are 2, 4, 8, 16, 32, 64, 128
  and 256\,mJy\,arcsec$^{-2}$. Note the strength of the continuum
  emission of the unresolved nucleus, and the ghost effect which can
  be seen to the south.  b) An ISOCAM image of the circumnuclear
  star-forming regions, obtained by mapping the UIB feature at
  7.7\,$\mu$m after the removal of the underlying continuum with
  contours at the same wavelength. The contour levels are 0.75, 1,
  1.25, 1.5, 1.75, 2, 2.25, 2.50, 2.75, 3 and
  3.25\,mJy\,arcsec$^{-2}$. We indicate the location of the nucleus
  with a cross ($+$), as well as the position angle of the stellar and
  gaseous bar ({\it solid line}).  c) A SCUBA map at 450\,$\mu$m of
  the cold dust distribution (\cite{Papadopoulos99}) with contours of
  the MIR image presented in (b).  d) Same ISOCAM contours as in (b),
  superimposed on the molecular gas distribution traced by the
  {$^{12}$CO(1-0)} emission line convolved to the ISOCAM spatial
  resolution of $\sim$5\,{\arcsec} (Schinnerer et al. 2000). The FWHM
  of the PSF for each image as well as the image scale indicated by a
  bar of 720\,pc (10\,{\arcsec}) in length, are respectively included at
  the right and left corners at the bottom of each panel.}
\label{fig:map_star}
\end{figure*}

Several ghost images produced by multiple reflections between the
detector and the filter plane are particularly present in CVF
observations containing a bright point-like source (\cite{Okumura}
1998).  The primary ghost which results from the first reflection is
clearly observed in our CVF images (see
Figure~\ref{fig:map_star}a). Its overall shape is characterized by a
ring structure corresponding to the defocussed pupil
image. Nevertheless, the surface brightness of the most luminous ghost
is weak and represents only $\sim$7$\%$ of the average surface
brightness of the central region (40\,{\arcsec} in
diameter). Moreover, as seen in Figure\,5--16 in \cite{Okumura} (1998)
which presents a ghost pattern similar to that of our observation, the
ghost image is located essentially outside of the strong point-like
source. Thus we conclude that the ghost did not affect significantly
the photometry of the central region and therefore did not introduce a
systematic bias in our analysis of the data.

The surface brightness of the PSF of the nucleus averaged over a
region of 15\,{\arcsec}$\times$15\,{\arcsec} is one order of magnitude
greater than the surrounding starburst regions (see
Figure~\ref{fig:map_star}a), so special care had to be taken in order
to study the MIR spectral properties of the whole galaxy, and
to correctly interpret the morphology of the environment
surrounding the nucleus.  The first method used was to subtract the
contribution of the bright unresolved nucleus. This was done using a
set of experimental PSFs created by observing stars during the
calibration phase of the camera.  An iterative method, which selected
the best available PSF fit, was applied and the nuclear emission was
subsequently removed. This subtraction of the central unresolved point
source revealed two areas of strong MIR emission located in the
circumnuclear ring which correspond to massive star-forming regions.
Extensive tests showed that the use of experimental PSFs resulted in
images with no observable artifacts up to 14\,$\mu$m. Beyond this
limit, proper removal of the nuclear contribution was problematic
since the experimental PSFs were obtained with various broad-band
filters, and their large profiles beyond 14\,$\mu$m could not
reproduce anymore that of the AGN observed with the CVF. The use of
theoretically calculated PSFs was also envisionned but the results
were of lower quality and consequently this method was not followed.

A second approach was to isolate the emission from MIR spectral
features such as the various Unidentified Infrared Bands (UIBs) and to
produce images of the spatial distribution of those features across
the galaxy. This was done by identifying the spectral band of
interest, which for the brighter UIB at 7.7\,$\mu$m was between 7.3
and 8.3\,$\mu$m, in all observed positions (pixels) on the galaxy and
subtracting the underlying continuum emission (see
Figure~\ref{fig:spec_star} for an illustration of the method). We have
computed the flux density between 7.3 and 8.3\,$\mu$m in order to
facilitate the comparison between the surface brightness of the UIB
map and the original map including the continuum (see
Figures~\ref{fig:map_star}a and \ref{fig:map_star}b).  Since it has
now been widely demonstrated (see e.g. \cite{Laurent} 2000a, and
references therein) that UIB emission is preferentially associated
with regions of star formation and is markedly depleted in regions
where a very hard radiation field is present such as those near an
AGN, the spatial distribution of the UIB emission should trace the
starburst regions of NGC\,1068 quite accurately. Indeed, a picture 
similar to the one given by the PSF subtraction was obtained with this
method, providing confidence on the reliability of the
results. Consequently, this mapping of the spatial distribution of the
UIB emission was used to produce the final picture of the starburst
regions (cf. Figure~\ref{fig:map_star}b) and to analyze their
morphology. In the central region
(15\,{\arcsec}$\times$15\,{\arcsec}), the flux variation between each
individual pixel resulting from the non-perfect correction of the
jitter effect creates artificial noise and prevents us from estimating
accurately the UIB emission. Furthermore, the [NeVI] emission line at
7.7\,$\mu$m does contaminate the UIB emission at that wavelength 
(\cite{Lutz2000}). To
address this problem in a region of 15\,{\arcsec}$\times$15\,{\arcsec}
around the nucleus, where no UIBs were detected, we used as a template
the starburst spectrum of Figure~\ref{fig:spec_star} in order to
estimate an upper limit for the 7.7\,$\mu$m UIB emission based on the
detectability of the 6.2$\mu$m feature. The structures we detect
applying this method are consistent with the 10.8\,$\mu$m observations
of \cite{Telesco} (1988).  In particular, the agreement between the
position angles of the different MIR spots suggests that the
orientation angle of the ISOCAM data is well defined.  This is also
supported by the fact that no such angular uncertainties were ever
reported in ISOCAM deep surveys where they would have been easily
observed, because of the large number of detections on large spatial
scales.

Due to their good spatial resolution, our observations also allowed us
to analyze the different spectral energy distributions (SEDs) of the
nuclear environment as well as the MIR emission in the ring
surrounding it.  Whereas the spectrum of the star-forming regions was
obtained after the removal of the contribution from the unresolved
point source, via the PSF subtraction method described earlier, the
spectrum of the nucleus was directly measured in the CVF data and
estimated over a central region of 700\,pc in diameter. The only
correction applied was to account for the varying size of the PSF at
different wavelengths. The ISO-SWS spectrum of the nuclear region over
an aperture of 14\,{\arcsec}$\times$20\,{\arcsec} for 2.5--12\,$\mu$m
and 14\,{\arcsec}$\times$27\,{\arcsec} for 12--27.5\,$\mu$m presented
by Lutz et al. (2000) is similar in shape and intensity to the ISOCAM
spectrum of the AGN in Figure~\ref{fig:agn_prop}, while the intensity
offset between the two spectra is less than 20$\%$.  Our observations
are also consistent to a 20$\%$ level with the 8--13\,$\mu$m spectrum
published by \cite{Roche1984}. Ground based observations with broad
band filters obtained by Rieke $\&$ Low (1975) and more
recently by \cite{Bock1} (2000) show also a general agreement with our
data since the differences are lower than 30$\%$ and 15$\%$
respectively.

The relative photometric uncertainty on the AGN spectrum is mainly
produced by 1) the non-perfect correction of the detector memory
effect, estimated at 10$\%$ and 2) by the error on the PSF aperture
correction which can reach 10$\%$ at longer wavelengths. The relative
uncertainty varies from 10$\%$ at 5\,$\mu$m up to 15$\%$ at
16\,$\mu$m.  We estimate that the absolute uncertainty of our
photometric measurements is \,20$\%$ for the AGN spectrum. This value
is fully consistent with the different MIR observations of the central
region of NGC\,1068 obtained using a large variety of instruments.
For the starburst spectrum presented in Figure~\ref{fig:spec_star},
the photometry can be strongly affected by a residual AGN contribution
at wavelengths above 14\,$\mu$m. Nevertheless, the shape of the UIBs is
very similar to that observed in other starburst galaxies
(\cite{Laurent} 2000a).  Based on our experience with ISOCAM data we
estimate that an uncertainty of 30\,$\%$ for the absolute photometry
constitutes a conservative upper limit for the starburst contribution,
which is typical for well detected extended sources.

\section{The central region of NGC\,1068 in the Mid-Infrared}  
\label{sec:analysis}  
     
\subsection{Morphology of the circumnuclear environment}  
\label{subsec:morpho}  
  
The images obtained by mapping the different UIB features at 6.2, 7.7,
8.6 and 11.3\,$\mu$m are consistent with one another. In
Figure~\ref{fig:map_star}b, we map the spatial distribution of the
7.7\,$\mu$m feature, which is one of the strongest UIBs. One of the
main advantages in using the 7.7\,$\mu$m feature is that contrary to
the UIBs at 8.6 and 11.3\,$\mu$m, its intensity is not affected by the
silicate absorption at 9.7\,$\mu$m (see the MIR spectrum of Arp\,220
in \cite{Vassilis}). It is also preferred to the feature at
6.2\,$\mu$m because of its higher flux, providing a better signal to
noise ratio when subtracting the underlying continuum emission.  As we
discuss in the following section, the 7.7\,$\mu$m feature is
considered as a powerful indicator of star forming environments
(e.g. \cite{Genzel} 1998, \cite{Rigopoulou}). The two starburst
regions detected with ISOCAM are located North-East and South-West of
the nucleus, in agreement with the spots of MIR emission observed at
10.8\,$\mu$m by \cite{Telesco} (1988). They are roughly extending on
both sides of the extremities of the stellar bar ($R \sim
1.5\,\mbox{kpc}$, P.A. $\sim 48\degr$, \cite{Scoville}). This
concentration near the ends of the bar may be the result of density
waves that compress and shock the gas present in the spiral structure,
which in turn triggers the star-forming activity (\cite{Telesco} 1988,
\cite{Tacconi}, \cite{Eva}).

While the MIR observations allow us to map the hot dust and the UIB
emission, the FIR emission reveals the distribution of the cold dust
in the galaxy. FIR observations of NGC\,1068 show that, in general,
the cold dust is spatially correlated with the emission of the CO/HI
gas (\cite{Papadopoulos99}). The extended FIR emission, which
is attributed to a dust component with temperature of
T$\sim$10\,K, seems to be associated with regions of high HI column
density. More concentrated and warmer dust (T$\sim$30\,K) though, is
found in the inner starburst and it follows the distribution of the
molecular gas.  In Figure~\ref{fig:map_star}c, the ISOCAM 7.7\,$\mu$m
image is overlaid on the cold dust distribution traced by the
450\,$\mu$m FIR image obtained with SCUBA (\cite{Papadopoulos99}). The
two images have been translated with respect to each other, so that
the positions of the unresolved nucleus coincide.  Comparing the
ISOCAM and SCUBA data, a striking shift between the dominant peaks of
emission in the Mid and Far-Infrared is apparent. This shift can not
be a residual in the MIR data due to the contamination from the AGN
continuum since the same morphology is also observed in the other UIB
maps at 6.2, 8.6 and 11.3\,$\mu$m, even though the point-like emission
of the AGN is different at these different wavelengths.  A systematic
error in the orientation angles of either the ISOCAM or SCUBA maps
given the observed performance of both instruments over the years is
highly unlikely. Furthermore, it does not seem possible to
exactly overlay all the features of both maps with a simple
rotation.

Since the inner regions of NGC\,1068 rotate counterclockwise,
the shift may indicate that the main features of the
circumnuclear ring detected with ISOCAM lead by $\sim$\,5\,{\arcsec}
($\sim$\,360\,pc) those observed by SCUBA.  We mentioned already that
the spots of emission revealed at submillimeter wavelengths could be
attributed to large amounts of cold dust heated by the local
interstellar radiation field. On the contrary, the UIB emission traces
recent star-forming regions located at the extremities of the bar.
This type of ISM morphology and formation of shocks in spiral arms
have been predicted by theoretical models (e.g. \cite{Athanassoula}),
where one observes that density enhancements in the gas
coincide with the leading edges of rotating primary bars.  Strong UIB
emission along the leading edges of barred structures has also been
observed by ISOCAM in other galaxies such as NGC\,1097
(\cite{Roussel}) and Centaurus\,A (\cite{Mirabel}). Finally, we note
that the UIB emission observed in NGC\,1068 correlates well with the
star formation spots revealed in H$\alpha$ (\cite{BHSa}, \cite{BHSb}),
as well as the Br$\gamma$ emission which also peaks at the extremities
of the stellar bar (\cite{Davies}).

The comparison between the UIB emission detected by ISOCAM and the
distribution of molecular gas obtained with the IRAM interferometer
(\cite{Eva}) is presented in Figure~\ref{fig:map_star}d.  The
{$^{12}$CO(1-0)} emission clearly resolves the central region into two
spiral arms extending from a molecular bar.  In the South-Western
knot, the molecular gas correlates well with the cold dust (see
Figure~2 of \cite{Papadopoulos99}), and therefore slightly follows
behind the MIR peak observed with ISOCAM.  In the northern spiral
structure, we also observe the same configuration with the
MIR emission remaining enclosed near the extremity of the bar whereas
the molecular gas and the cold dust are more extended along the arm.
This implies that a large fraction of the molecular gas available for
the star-formation process has not yet been converted into stars. The
shift between the MIR and CO/FIR emission observed in both spiral arms
may indicate, again, that the formation of young stars which
heat the UIB emitting grains requires the presence of shocks and
density enhancements rather than merely large amounts of
gas. It is unclear from our data whether a bar of hot dust is
present since the PSF of the AGN contaminates the central field of
view where such a bar would be expected. Nevertheless, the comparison
with the CO and FIR observations presented earlier, suggests
that a possible hot dust bar could lead the stellar/FIR bar and
it would appear at the place where shocks are being formed in the
leading edge of the gaseous bar.

\subsection{Spectral properties}  
\label{subsec:spectro}  
  
\subsubsection{The starburst energy distribution}  
\label{subsub:starburst}  

\begin{figure}  
\resizebox{\hsize}{!}{\includegraphics{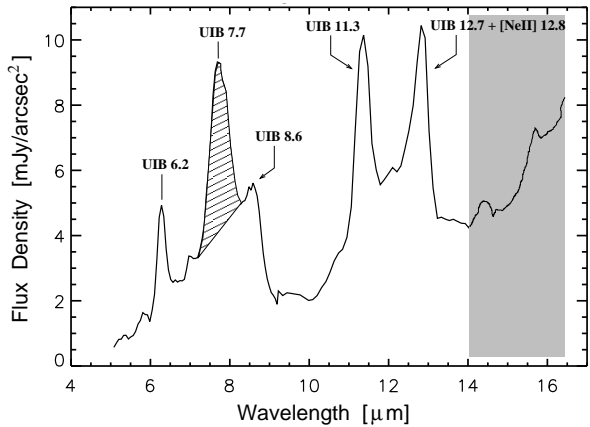}}  
  \caption{Spectrum of the starburst regions after removing the AGN
  contribution. It has been obtained over a region of 15\,{\arcsec} in
  diameter, centered on the MIR peak of emission located South-West of
  the nucleus ($\alpha$=02h42m39.9s,
  $\delta$=-00$^{\circ}$00$'$55.3$''$, J2000) and it is very similar
  to the MIR spectra of the other circumnuclear star forming regions
  of NGC\,1068. A proper removal of the AGN contribution could not be
  performed beyond 14\,$\mu$m (grey part of the plot, see details in
  section~\ref{sec:red}). The flux of the UIB at 7.7\,$\mu$m is
  estimated by integrating the hatched area between 7.3 and
  8.3\,$\mu$m.}
\label{fig:spec_star}
\end{figure}  
 
The MIR SEDs of the circumnuclear starburst regions show no spatial
variation and the spectrum of the South-West peak is presented in
Figure\,\ref{fig:spec_star}. It is typical of MIR spectra detected in
star forming regions and can be decomposed into two main components
(e.g. \cite{Tran}, \cite{Laurent} 2000a):

1) a very steeply rising continuum as a function of wavelength, which
becomes important at 10--16\,$\mu$m. This is the most prominent
characteristic of MIR spectra observed in galaxies showing evidence of
intense starburst activity (\cite{Laurent} 2000a).  It is attributed
to the Very Small Grains (VSG, \cite{Desert}) and is produced in
\ion{H}{ii} regions, where dust is heated by massive hot young
stars. It is thus considered as a good indicator to trace regions of
massive star formation activity usually obscured in the visible
(\cite{n4038}). An example of a MIR spectrum from a pure HII
Galactic region can be seen in \cite{Cesarskyb}.
 
2) the family of the  Unidentified Infrared Bands, which are  centered
at 6.2, 7.7, 8.6,  11.3 and 12.7\,$\mu$m,  and are clearly visible in
our  spectra.  They are attributed  to  C=C  and  C\---H vibrations in
Polycyclic Aromatic Hydrocarbon molecules (PAH, \cite{LegerPuget},
\cite{Alamandola}). These molecules are transiently heated by 
the stellar radiation field of massive stars and they are found in
photo-dissociation regions (PDRs) surrounding HII regions.  As a
result, these UIB features are associated with star formation activity
and they are typically found in MIR spectra of galaxies which form
stars in either quiescent or more active manner (\cite{Helou} 2000).

Often, several ionic emission lines are also observed in the MIR
spectra.  The most ubiquitous one is the 12.8\,$\mu$m [NeII] line ($Ep
= 22\,\mbox{eV}$) which is present in almost all spectra of star
forming regions, even though at the spectral resolution of ISOCAM
($\lambda / \Delta\lambda = 40$), it is blended with a UIB feature at
12.7\,$\mu$m.  It has been noted that in most ISOCAM spectra
of regions with low level of star formation activity, the
intensity in the 12.7\,$\mu$m feature is smaller than the one at
11.3\,$\mu$m (\cite{Boulanger}, \cite{Cesarskyc}).  The fact
that in NGC\,1068 we observe the opposite is a strong indication that
despite the blending the ionic line provides most of the flux.
Another line, detected at 7\,$\mu$m, is attributed to the emission of
the [ArII] line at 6.99\,$\mu$m which has been clearly
identified in starburst galaxies with ISO-SWS (\cite{Sturm},
\cite{Nat}). It has a rather low ionizing potential
($Ep\,=\,16\,\mbox{eV}$) and as a consequence is produced essentially
in starburst environments. The [ArII] line is seen only at
weak levels in NGC\,1068 by Lutz et al. (2000), possibly
because of the dilution of the starburst spectral signature by the
AGN, due to the extended size of the SWS aperture.

\subsubsection{The nuclear Mid-Infrared spectrum}  
\label{subsub:AGN}  
  
The 5--16\,$\mu$m spectrum of the nuclear region is displayed in
Figure~\ref{fig:agn_prop}. The difference, compared to the MIR SED
found in the starburst regions, is striking and its overall spectral
shape is consistent with that found in other AGNs observed in the MIR
(\cite{Lutz}). It is characterized by a strong continuum commonly
attributed to the very hot dust which is present in the torus of
molecular gas surrounding the nucleus as described in the unified
scheme (see Krolik 1999, and references therein). Similarly rising
continua can also be observed in nearby resolved HII regions
(\cite{Alessandra}). The MIR spectra of AGNs though are much flatter,
and show an important continuum emission even at short wavelengths
(3--6\,$\mu$m), because of dust particles heated to nearly
their evaporation temperature (T\,$\sim$\,1000\,$\mbox{K}$ for
silicates and T\,$\sim$\,1500\,$\mbox{K}$ for graphites) by the
AGN. Since we know that NGC\,1068 hosts an AGN, our spectrum thus
confirms that a telltale sign of its presence is the high continuum
level between 5 and 6\,$\mu$m (\cite{Laurent} 2000a).

Highly ionized emission lines such as [NeV] at 14.3\,$\mu$m
($Ep\,=\,97\,\mbox{eV}$) and [NeVI] at 7.6\,$\mu$m
($Ep\,=\,126\,\mbox{eV}$) are seen in our spectrum. These lines are
clearly visible in the high spectral resolution spectrum obtained by
ISO-SWS (\cite{Lutz2000}), which provides confidence that our
detections are not due to other faint UIBs, often observed very close
to these forbidden lines (\cite{Nat}).  Although the [NeV] and
[NeVI] emission lines are also observed in supernova remnants (SNRs),
the very weak MIR continuum associated with SNRs (\cite{Oliva}) can
not account for the MIR spectrum of the nucleus of NGC\,1068. The
[NeIII] line is also clearly visible at 15.6\,$\mu$m,
contrary to [NeII] at 12.8\,$\mu$m which is hardly detected. An effort
was made to measure the [NeIII]/[NeII] ratio, which is often used to
quantify the strength of a radiative field. However, subtracting the
underlying continuum was not straightforward and only an upper limit
could be estimated for the [NeII] intensity. We obtained a rather high
[NeIII]/[NeII] ratio ($\gtapp 2.5$), compared to what has been
generally found in starburst galaxies ([NeIII]/[NeII] $\ltapp 1$,
\cite{Thornley}). This ratio is consistent with the value of 2.3
obtained with the ISO-SWS spectroscopic observations (\cite{Lutz2000})
and argues for a harder radiation field in the nucleus.

Other than the continuum and the Neon forbidden lines, we also note
the absence of the family of UIBs in the nuclear MIR spectrum.  This
depletion is often interpreted as the destruction of PAH molecules by
the extremely intense X-ray radiation field of the AGN (\cite{Voit91}, 
\cite{Allain}). In
principle, a comparison between the surface brightness of the
starburst and the AGN can be carried out in order to estimate the
level of the UIB contribution in the nuclear spectrum.  However, as
seen earlier, the noise associated with the AGN continuum is already
higher than the average surface brightness of the star formation
regions observed in the disk around the circumnuclear starburst (see
Figure~\ref{fig:map_star}b). This prevents us from examining whether
the lack of UIBs in the AGN spectrum is really due to a physical
absence of the PAHs, or rather to a very low emission from the UIBs in
the central regions which could originate from 1) foreground emission
from the disk, 2) weak star formation or 3) the AGN. Nevertheless,
even if the UIBs are present in the AGN spectrum at a faint
level, their intensity is much lower than the UIB emission
from the circumnuclear starburst presented in
Figure~\ref{fig:map_star}b.  Measuring the upper limit for UIB
emission from a region of 15\,{\arcsec}$\times$15\,{\arcsec}
centered on the nucleus of the galaxy, we estimate that at least
$\sim$\,95\,$\%$ of the UIB emission of NGC\,1068 originates outside
the nucleus.  Thus the UIB features provide a reliable tracer of star
formation unrelated to the AGN activity. This issue, regarding the
possible destruction of UIBs in the central regions of the AGNs could
be addressed in the future using more sensitive infrared
spectrographs, such as the IRS on board SIRTF (\cite{Houck}).

\section{Discussion}
\label{sec:discuss}

\subsection{NGC\,1068: A key object to distinguish an AGN from a starburst}

The proximity of NGC\,1068 coupled with the good spatial and spectral
resolution of our data allows us to clearly disentangle both the
spatial distribution and the spectral properties of the AGN from the
emission of the surrounding starburst regions. Similar
differences in spectral characteristics have already been reported in
other nearby prototypes such as Centaurus\,A (\cite{Mirabel}) which,
at a distance of 3.5\,Mpc, is the closest radio galaxy to Earth. Such
objects may be used as templates to distinguish AGN from starburst
emission and evaluate the fraction of their respective
contribution in spectra of more distant galaxies where the angular
resolution is not sufficient to spatially separate the sources
of the two components.  Diagnostics based on this approach have been
developed recently in the MIR by \cite{Genzel} (1998) and
\cite{Laurent} (2000a).  As we have already mentioned, the striking
difference of the relative strength of UIBs observed in various
radiation fields appears to be a canonical feature in the MIR SED in
nearby objects: UIB emission is not detected in AGNs whereas it is
very pronounced in regions of massive star formation. This is clearly
demonstrated by the very distinct spatial separation between
starburst regions and AGN visible in our 7.7\,$\mu$m map of NGC\,1068
(see Figure~\ref{fig:map_star}).

\begin{figure}    
\resizebox{\hsize}{!}{\includegraphics{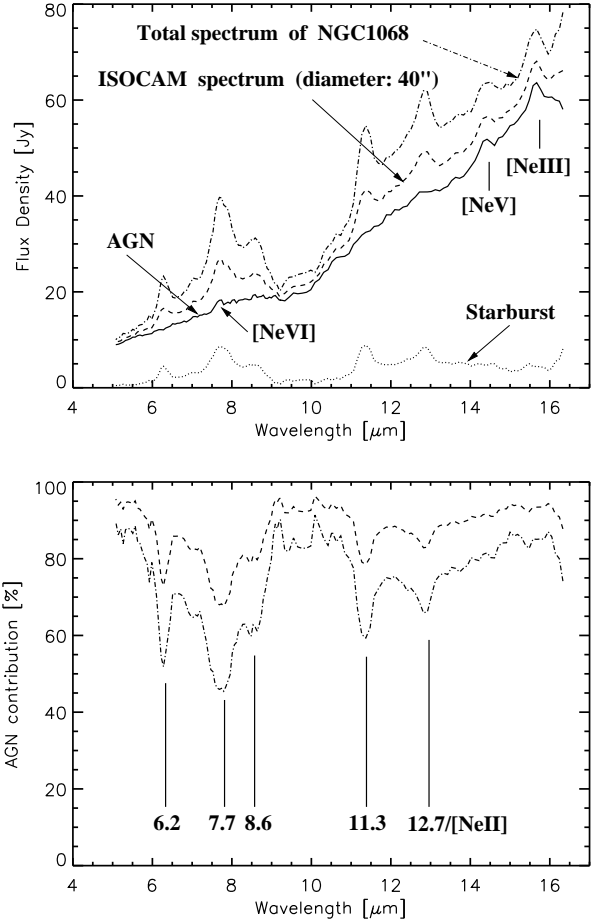}}  
  \caption{{\it Top:} Decomposition of the total MIR spectrum of
  NGC\,1068 integrated over a region 40\,{\arcsec} in diameter ({\it
  dashed line}) into the emission from the nucleus, which is dominated
  by the AGN continuum ({\it solid line}) and the one originating from
  the circumnuclear starburst regions ({\it dotted line}). An estimate
  of the total integrated emission of NGC\,1068 based on a comparison
  with B-band photometry ({\it dashed-dotted line}, cf. section
  4.1). {\it Bottom:} Fraction, in percent, of the unresolved nuclear
  flux to the ISOCAM MIR emission within 40\,{\arcsec} in diameter
  ({\it dashed line}) and to the total MIR emission ({\it
  dashed-dotted line}).}
\label{fig:agn_prop}
\end{figure}

The spatial resolution of ISOCAM does not allow us to study the
distribution of the MIR emission near the extended NLR observed with
high resolution MIR imaging (\cite{Braatz}, \cite{Bock0}, \cite{Bock1}
2000, \cite{Alloin}).  Nevertheless, those high resolution images have
shown that nearly 70$\%$ of the MIR flux originates from within a
region of 1\,{\arcsec} around the nucleus, which spatially
coincides with the ionization cone. This implies that the integrated
nuclear emission observed by ISOCAM in Figure~\ref{fig:agn_prop} is
produced essentially by hot dust located in the ionization
cone and not in the torus.  Hence one would expect that UIBs are also
absent in those regions extending up to 70\,pc (1\,{\arcsec}) from the
nucleus. Moreover, the MIR SEDs of these regions rise from 7.9
to 10\,$\mu$m (\cite{Bock1} 2000). This further suggests that in the
range of 6--9\,$\mu$m no strong UIBs are present, since if they were
the spectrum would have been flatter. Consequently, the radiation
field produced by the AGN must be able to heat the dust at
T$\gtapp$150\,K (\cite{Lumsden}) and to destroy UIBs at distances
of several tens of parsecs from the nucleus as predicted by
Voit (1992).  Recently, MIR observations at 11 and 20\,$\mu$m of the
central 6\,{\arcsec}$\times$6\,{\arcsec} region of NGC\,1068 revealed
even more extended emission 4\,{\arcsec} away from the AGN which
follows the radio jet structure (\cite{Alloin}). This resolved MIR
emission which extends up to $\sim$\,300\,pc from the central region
including the NLR and the torus (0.6\,{\arcsec}$\times$0.9\,{\arcsec})
contributes about 5$\%$ of the total MIR emission at 11.2\,$\mu$m and
should contribute to the ISOCAM spectrum of the AGN, but only at faint
levels. The observed MIR spectral differences between star forming
regions and the NLR could be used in combination with the NLR emission
line ratio proposed by Genzel et al.(1998) to identify AGNs with
optically thick tori in MIR.

Another noticeable feature in the nuclear spectrum of NGC\,1068
spectrum is the weak silicate absorption at 9.7\,$\mu$m. Based on a
screen model (dust absorption law of \cite{Dudley}) applied to a power
law of spectral index $\alpha=-1.7$, the best fit of the nuclear
spectrum leads to a visual extinction of only $A_v=7$\,mag (see also
Figure\,10 of \cite{Laurent} 2000a). This is consistent with the value
($\tau_{9.7}=0.51$; $A_v\sim 7.5$\,mag) inferred by \cite{Roche1984}.
One may wonder why the silicate absorption at 9.7\,$\mu$m is fairly
weak in our spectrum ($A_v=7$\,mag) which is in rather sharp contrast
with the strong extinction, $A_v \sim 40$\,mag, derived for the very
hot dust ($\sim$\,1000\,K) seen through the obscuring torus using
polarimetry techniques on NIR emission lines (\cite{Young95}) and on
Near and Mid-Infrared imaging (\cite{Lumsden}).  This apparent
contradiction already mentioned by several authors
(e.g. \cite{Braatz}, \cite{Efstathiou}, \cite{Bock0}) can be explained
by the fact that we observe an extended MIR emission. High
resolution images have shown that $\sim70\%$ of the MIR flux
originates from optically thin dust in this region which is
unaffected by extinction and also dilutes the more obscured MIR
emission from the torus. Recently, \cite{Bock1} (2000) have clearly
confirmed the absence of silicate absorption in the regions of
the ionization cone regions whereas the unresolved nuclear region
presents a clear dip at 10\,$\mu$m. Furthermore, Maiolino et
al. (2000) have shown observational evidence that dust in the
environment of the AGN should have different properties from the
dust found in the Galactic diffuse interstellar medium. This
dust would be characterized by a deficiency of small dust grains less
than 3\,$\mu$m in size,
such as the silicate particles responsible for the absorption
feature at 9.7\,$\mu$m. The weakness of silicate absorption could then
be explained in terms of both a small intrinsic absorption in the
ionization cone, from where a large fraction of the MIR
emission originates, as well as a smaller fraction of silicates in
the dusty torus.

Since in NGC 1068, our data allow the separation of the
starburst and AGN components, we can investigate the
reliability of MIR activity diagnostic tools as a function of
the amount of extinction affecting the AGN component. To do so, we
apply an increasing extinction (in a screen model and assuming a
standard Galactic extinction law) to the AGN component before
combining it to the starburst component. As the extinction
increases, one notes that not only does the silicate feature at
9.7\,$\mu$m become deeper, but also the continuum at shorter
wavelengths (5--10\,$\mu$m) decreases significantly, becoming less
important than the UIB emission (\cite{Laurent2}). According to the
recent discovery by Maiolino et al. (2000), the depletion of silicate
particles in the AGN torus will produce the same extinction
to the continuum but without a deep silicate absorption
feature. Since, as we mentioned in the previous section, this hot
continuum is used as an AGN indicator, the total spectrum of the
galaxy integrated over a region 40\,{\arcsec} in diameter would show
prominent UIB features, leading to an ``apparent''
starburst-classified galaxy. We note however that in the case
of NGC\,1068, the additional extinction should be applied only to the
torus emission which contributes only 30\% to the MIR nuclear
emission.  Therefore in this galaxy the hot dust emission would never
be completely suppressed by a strong absorption in the torus.

A point which should be noted is that only the inner part of the
galaxy (96\,{\arcsec}$\times$96\,{\arcsec}) has been observed with
ISOCAM and that the MIR contribution of the outer disk is not taken
properly into account (see Figure~\ref{fig:map_view}). If one were to
use a B-band image of the galaxy (\cite{schild}) as a tracer of star
formation in the galaxy, nearly 60\% of it occurs outside the area of
40\,{\arcsec} in diameter, where the strong circumnuclear starburst
revealed by the MIR band emission and the CO barred spiral pattern are
found. Even though the radial scale length of the dust should
be smaller than that of the blue light, since the MIR emission of the
disk is very likely UIB rich (see Roussel et al. 1999, 2000), one
would expect that the starburst characteristics in the MIR will
increase if we take the whole galaxy into account. This is an
important effect which should be considered when one extrapolates
these results to distant galaxies (\cite{Moriond}). Only 20\%
of the total blue light is emitted from regions outside the area
mapped by ISOCAM. Using the UIB map (Figure~2b) we compute that the
fraction of 7.7\,$\mu$m emission from the region outside the inner
starburst ring (40\,{\arcsec} in diameter) compared to the total
integrated UIB emission on the ISOCAM field of view is
53$\%$. A photometric calculation over the same areas on a
B-band image gives 55$\%$ suggesting that the B-band light and the
UIB emission have similar scale lengths in the disk of NGC 1068.
Consequently we can use the blue light and the high signal to noise
UIB measurements within the inner 40\,{\arcsec} of the galaxy to apply
a correction for the unaccounted 60\% of the MIR flux from the
spatially extented UIB emission (see Figure~\ref{fig:agn_prop}). Doing
so, we have implicitly assumed that the extended disk emission of
quiescent star formation does not differ strongly from the starburst
emission along the ring observed with ISOCAM. Helou et al. (2000) have
recently shown that among 28 galaxies with diverse properties
(R(60/100)$\sim$0.28-0.88), the MIR spectral shape from 5.7 to
11.6\,$\mu$m varies only weakly. Hence, our proportional correction
factor applied to the starburst spectrum should give us a consistent
result in the UIB region under 12\,$\mu$m even though some
uncertainties are still present at longer wavelengths
(12--16\,$\mu$m). One can further check the consistency of this
photometric correction for the extended size of NGC\,1068, by
comparing the observed IRAS 12$\mu$m flux density of the
galaxy to the one we can calculate using the ISOCAM SED and the
transmission of the IRAS 12$\mu$m filter. The value we find using our
CVF data is 36.8\,Jy for the integrated emission before correction
(40\,{\arcsec} in diameter), and 43.6\,Jy for the total integrated
emission after correction (200\,{\arcsec} in diameter).  One notes
that even though the correction of the starburst UIB emission is
important (60$\%$), the overall MIR flux at 12$\mu$m is largely
dominated by the AGN and as result does not change by a large factor
(see Figure~\ref{fig:agn_prop}). The value we find for the equivalent
IRAS 12$\mu$m filter after the correction is slightly more
than the observed 12$\mu$m IRAS flux density, which was
quoted as 39.7\,mJy in Moshir et al. (1990) and 36.1\,Jy in Soifer et
al. (1989) assuming 10$\%$ of flux uncertainty. This suggests that the
correction of 60$\%$ (20\% of the MIR flux is outside of the ISOCAM
field of view) cannot be much larger at 8--15$\mu$m and therefore the
spectrum shown in Figure~\ref{fig:agn_prop} represents a good estimate
of the total MIR emission of NGC\,1068.

\subsection{AGN/starburst flux ratio}  
\label{subsec:propor}

In NGC\,1068, the total MIR spectrum integrated over a region of
40\,{\arcsec} in diameter has been decomposed into the contribution of
the unresolved AGN, and the residual which is attributed to the
circumnuclear starburst emission. In Figure~\ref{fig:agn_prop}, we
present the fraction of the MIR flux coming from the nuclear region
with respect to the total MIR emission detected from that galaxy as a
function of wavelength. One can notice that {\em the nuclear region
contributes $\sim85-95\,\%$ over most of the ISOCAM MIR spectrum} but
it decreases to $\sim70-80\,\%$ at the wavelengths corresponding to
the PAH emission (note how the 6.2, 7.7, 8.6, 11.3 and 12.7\,$\mu$m
features anti-correlate with the UIB emission of the startburst
spectrum of Figure~\ref{fig:spec_star}). Of course this fraction must
be viewed as an upper limit since the disk of NGC\,1068 is larger than
the field we imaged in the MIR. However, in the spectrum
corrected for this effect the AGN contribution still dominates the
total emission but at a lower level. The highest starburst
contribution to the continuum flux density (50\%) occurs at
7.7\,$\mu$m.  It is interesting to note that we can use this spectral
behaviour to disentangle the AGN from the starburst contribution using
only broad band imaging since the contrast of the AGN is enhanced at
5\,$\mu$m and 10\,$\mu$m compared to the starburst regions. Recently,
\cite{Krabbe} (2000) have shown that ground-based MIR observations at
10.5\,$\mu$m (N-band) are reliable in identifying obscured AGNs
previously only detected in hard X-rays (2-10\,keV). The N-band filter
is rather wide though (8 to 13\,$\mu$m) and integrates the UIBs at
8.6, 11.3 and part of the 7.7\,$\mu$m feature. If one is to
find obscured AGNs using their MIR spectral properties, then using the
M band filter (4.2-5.5\,$\mu$m) would be a better choice since at
these wavelengths both the UIB contamination is relatively faint and
the extinction caused by silicate particles is negligible.  This
dominance of the nucleus in the MIR contrasts strongly with what is
observed in the optical and FIR wavelengths. Using the SCUBA data, we
have estimated that the contribution of the nucleus does not represent
more than 25\% of the total emission at 450\,$\mu$m. Moreover, the
spatial distribution of the SCUBA image shows that the cold dust
emission is mainly associated with the starburst regions.

Finally, considerable efforts have been recently devoted to examine
whether luminous infrared galaxies (LIGs) are mainly powered by AGN or
starburst activity (see \cite{SandersMirabel} for a review). A recent
spectroscopic MIR survey (\cite{Lutz}) has shown that starbursts
dominate the energy budget in 85\% of the objects below
$3\times10^{12} L_{\sun}$, whereas AGNs seem to be always present in
ultraluminous galaxies with
$L_{IR}\,>\,3\times10^{12}\,L_{\sun}$. At $z=0.1$, 3\,kpc
corresponds to approximately 2\,{\arcsec} ($q_0\,=\,0.5,
H_0\,=\,75$\,km\,s$^{-1}$\,Mpc$^{-1}$), and represents the
typical scale of the circumnuclear starburst in NGC\,1068.  Thus, at
higher redshifts, the angular resolution of ISO (as well as the one of
SIRTF in the future) is not sufficient for a spatial AGN/starburst
discrimination. However, deep ISOCAM imaging surveys of distant
galaxies have revealed a significantly higher detection rate in the
LW3 band (15\,$\mu$m) than in LW2 (6.75\,$\mu$m) and the nature of
those objects resembles more that of starburst galaxies
(\cite{Aussel}, \cite{Elbaz}). It has been suggested though that this
result may also be due to an extreme absorption since recent X-ray
observations indicate that a large number of high redshift objects
harbor active nuclei (\cite{Mushotzky}). As such, our MIR spectral
imaging of the nearby prototypical Seyfert 2 galaxy NGC\,1068, may be
used as a template to investigate the observational bias on LIG deep
counts at cosmological distances.

\section{Conclusions}  
 
Based on the ISOCAM data presented in the previous sections we
conclude that:

1) The MIR emission of the Seyfert 2 galaxy NGC\,1068 is dominated at
$\sim$\,75\% by an unresolved point source, probably arising from hot
dust heated by the radiation field from the AGN. The MIR spectrum
associated with the AGN is characterized by a strong continuum and by
a non-detection of Unidentified Infrared Bands (UIBs). Recent MIR high
resolution imaging suggests that the intense and hard radiation from
the AGN is probably able to heat the dust and destroy the PAHs not
only in the inner torus but also at large distance in the narrow line
region. The AGN MIR spectrum also exhibits [NeV] and [NeVI] emission
lines, indicative of the hard radiation field.

2) More than 95\,$\%$ of the UIBs and at least 75\,$\%$ of the FIR
emission at 450\,$\mu$m originate from a circumnuclear starburst ring
of 3\,kpc diameter and from the disk of the galaxy.

3) The UIB map, which traces the young massive stars, shows a spatial
distribution which is in general agreement with the molecular gas and
the cold dust emission. However, the two peaks of UIB emission
observed in the circumnuclear arms of NGC\,1068 display a small
spatial shift of $\sim$\,5\,{\arcsec} ($\sim$\,360\,pc) towards the
leading edge of the arms at the extremity of the gaseous/stellar bar
where shocks and density enhancements have been predicted by
theoretical models of barred spiral galaxies.

\begin{acknowledgements}  
We are particularly grateful to L.~Tacconi, and P.~Papadopoulos, for
providing their published data displayed in Figure~\ref{fig:map_star},
and to the valuable comments of an anonymous referee.  IFM
acknowledges partial support from CONICET, Argentina.
\end{acknowledgements}

\end{document}